\documentclass[epj]{webofc}

\usepackage{graphicx}
\usepackage{bm}
\usepackage{epstopdf}

\woctitle{XLVI International Symposium on Multiparticle Dynamics}
\def\bea{\begin{eqnarray}}
\def\eea{\end{eqnarray}}

\def\pp{\mbox{$p$-$p$}}
\def\pa{\mbox{$p$-$A$}}

\def\auau{\mbox{Au-Au}}

\def\pbpb{\mbox{Pb-Pb}}
\def\aa{\mbox{$A$-$A$}}

\def\pt{$p_t$}
\def\yt{$y_t$}

\def\v2{$v_2$}

\begin{document}
\title{Rescuing the nonjet (NJ) azimuth quadrupole from the flow narrative}

\author{Thomas A. Trainor\inst{1}
}

\institute{CENPA 354290 University of Washington, Seattle, Washington, USA
          }

\abstract{According to the flow narrative commonly applied to high-energy nuclear collisions a cylindrical-quadrupole component of 1D azimuth angular correlations is conventionally denoted by quantity \v2\ and interpreted to represent elliptic flow. Jet angular correlations may also contribute to \v2\ data as ``nonflow'' depending on the method used to calculate \v2, but 2D graphical methods are available to insure accurate separation. The nonjet (NJ) quadrupole has various properties inconsistent with a flow interpretation, including the observation that NJ quadrupole centrality variation in A-A collisions has no relation to strongly-varying jet modification (``jet quenching'') in those collisions commonly attributed to jet interaction with a flowing dense medium. In this presentation I describe isolation of quadrupole spectra from pt-differential \v2(\pt) data from the RHIC and LHC. I demonstrate that quadrupole spectra have characteristics very different from the single-particle spectra for most hadrons, that quadrupole spectra indicate a common boosted hadron source for a small minority of hadrons that ``carry'' the NJ quadrupole structure, that the narrow source-boost distribution is characteristic of an expanding thin cylindrical shell (strongly contradicting hydro descriptions), and that in the boost frame a single universal quadrupole spectrum (L\'evy distribution) on transverse mass $m_t$ accurately describes data for several hadron species scaled according to their statistical-model abundances. The quadrupole spectrum shape changes very little from RHIC to LHC energies. Taken in combination those characteristics strongly suggest a unique {\em nonflow} (and nonjet) QCD mechanism for the NJ quadrupole conventionally represented by \v2.
}
\maketitle

\section{Introduction} \label{intro}

The flow narrative, believed by some to be of central importance to high-energy nuclear collisions~\cite{keystone,perfect},  is based primarily on $v_2$ data interpreted to represent elliptic flow -- azimuthal modulation of radial flow of a locally-thermalized bulk medium in non-central \aa\ collisions. Earlier versions referred to more-central \aa\ collisions at higher collision energies. More recently, ``collectivity'' in small systems (\pp, \pa) has been claimed as well, based on certain LHC data~\cite{cms,cmsridge}.
However, evidence strongly contradicting the flow narrative has accumulated over the past ten years: Differential analysis of \pt\ spectra from 200 GeV \auau\ collisions reveals no evidence for radial flow -- the blast-wave model said to measure radial flow responds instead to a predicted and observed strong jet contribution to spectra~\cite{hardspec}. Measurements of \pt-integral nonjet (NJ) $v_2$, based on model fits to 2D angular correlations that exclude a jet contribution (``nonflow'')~\cite{davidhq},  reveal $v_2$ systematics uncorrelated with ``jet quenching''~\cite{starraalim} contradicting claims for a dense bulk medium \cite{nohydro,noelliptic}. Equivalent systematic $v_2$ trends are observed for high-multiplicity \aa\ collisions and for \pp\ collisions down to negligible particle densities~\cite{ppquad}. There is no evidence for a QCD phase transition or changing equation of state.

In this talk I present {\em quadrupole spectra} inferred from $v_2(p_t,b)$ data that further contradict the flow narrative. Quadrupole spectra from recent LHC \pbpb\ data are compared to those from previous analysis of 200 GeV \auau\ data~\cite{quadspec}. Quadrupole spectra reveal a hadron source boost incompatible with Hubble expansion of a flowing bulk medium, and the spectrum shape is very different from the single-particle (SP) spectrum for most final-state hadrons.

\section{Quadrupole Spectrum Definition} \label{quaddef}

\pt-differential \v2\ is defined as the ratio of the quadrupole ($m = 2$) Fourier amplitude of the event-wise azimuth-dependent SP spectrum to the azimuth-averaged SP spectrum.
\bea \label{v2struct}
v_2(p_t,b) \hspace{-.05in} &=& \frac{V_2(p_t,b)}{\bar \rho_0(p_t,b)} =   \frac{V_2\{2D\}(p_t) + \text{jet contribution}}{(N_{part}/2)S_{NN}(p_t) + N_{bin}r_{AA}(p_t,b) H_{NN}(p_t)}.
\eea
As defined that ratio may include two jet contributions: (a) jet-related angular correlations in the numerator and (b) SP spectrum hard component $H_{AA}(p_t,b) \equiv r_{AA}(p_t,b) H_{NN}(p_t)$ in the denominator. The \v2\ definition assumes that almost all hadrons ``carry'' the quadrupole correlation component and therefore are described by the same SP spectrum that should then cancel in the \v2\ ratio. The NJ quadrupole Fourier amplitude, assuming source-boost azimuth distribution $\Delta y_{t}(\phi_r,b) = \Delta y_{t0}(b) + \Delta y_{t2}(b) \cos(2 \phi_r)$, can be expressed as~\cite{quadspec}
\bea \label{v2v2}
V_2(y_t,b)
&=&\frac{1}{2\pi} \int_{-\pi}^{\pi} d\phi\,  \bar \rho_0(y_t,b,\phi_r) \cos(2 \phi_r)
\approx
 p'_t\, \frac{ \Delta y_{t2}(b)}{2T_2} \, \bar \rho_2[y_t,b;\Delta y_{t0}(b)],
\eea
where $\phi_r$ is $\phi$ relative to a reference angle, $p_t'$ is \pt\ in the boost frame, transverse rapidity is $y_t \equiv \ln[(p_t + m_t)/m_h]$ and $\bar \rho_2[y_t,b;\Delta y_{t0}(b)]$ is the spectrum for those hadrons carrying the quadrupole correlation component, which may or may not be equivalent to $\bar \rho_0(y_t,b)$.

\section{Ideal Hydro} \label{ideal}

\begin{figure}[h]
\centering
  \includegraphics[width=.24\textwidth]{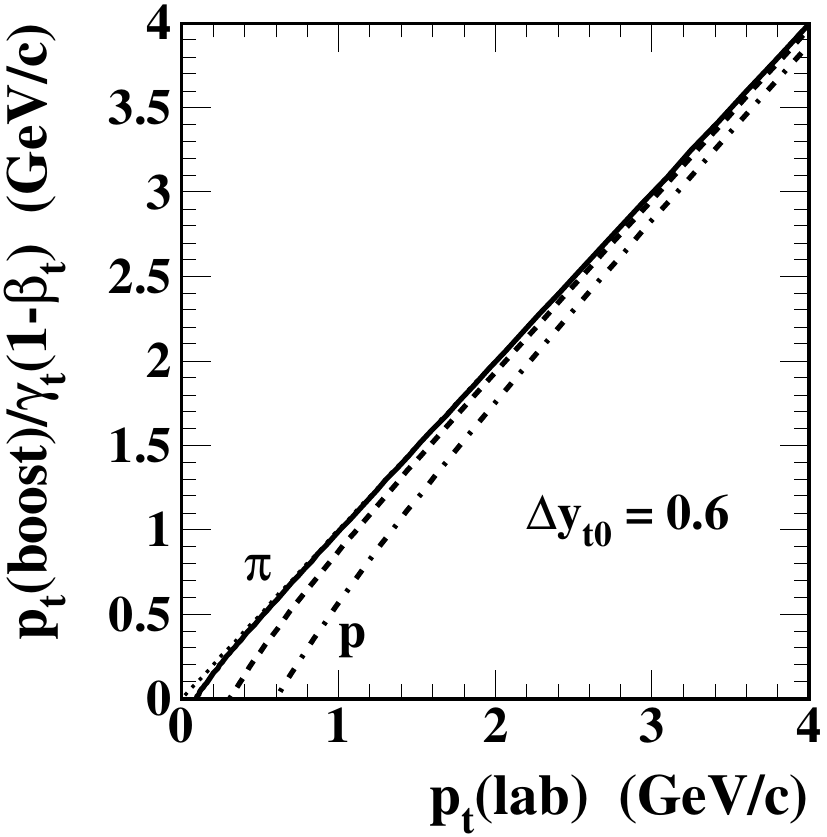}
\put(-70,70) {\bf (a)}
  \includegraphics[width=.24\textwidth]{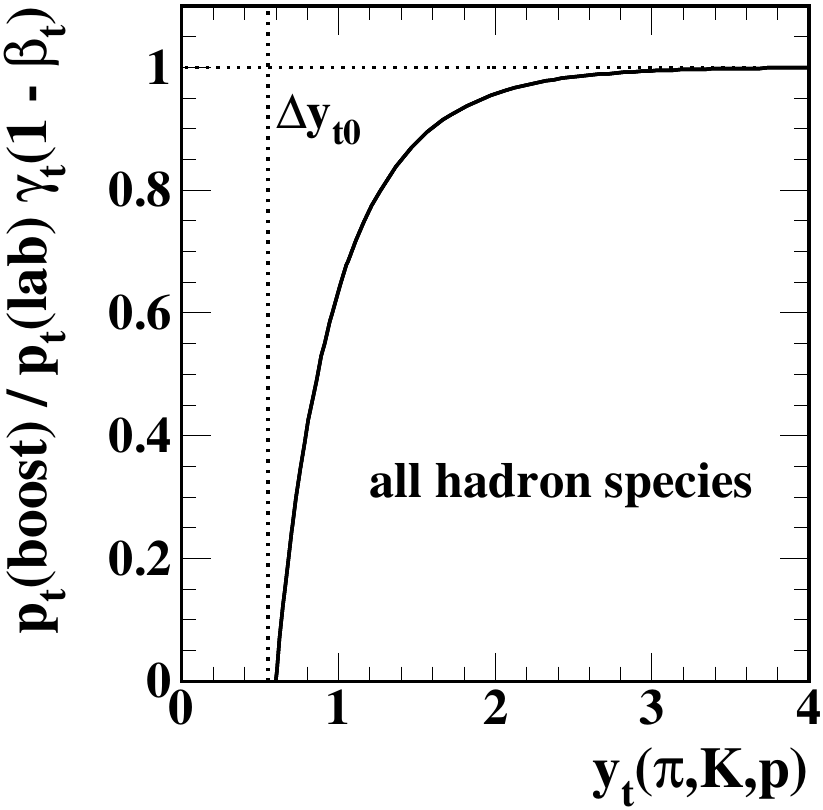}
\put(-23,70) {\bf (b)}
  \includegraphics[width=.24\textwidth]{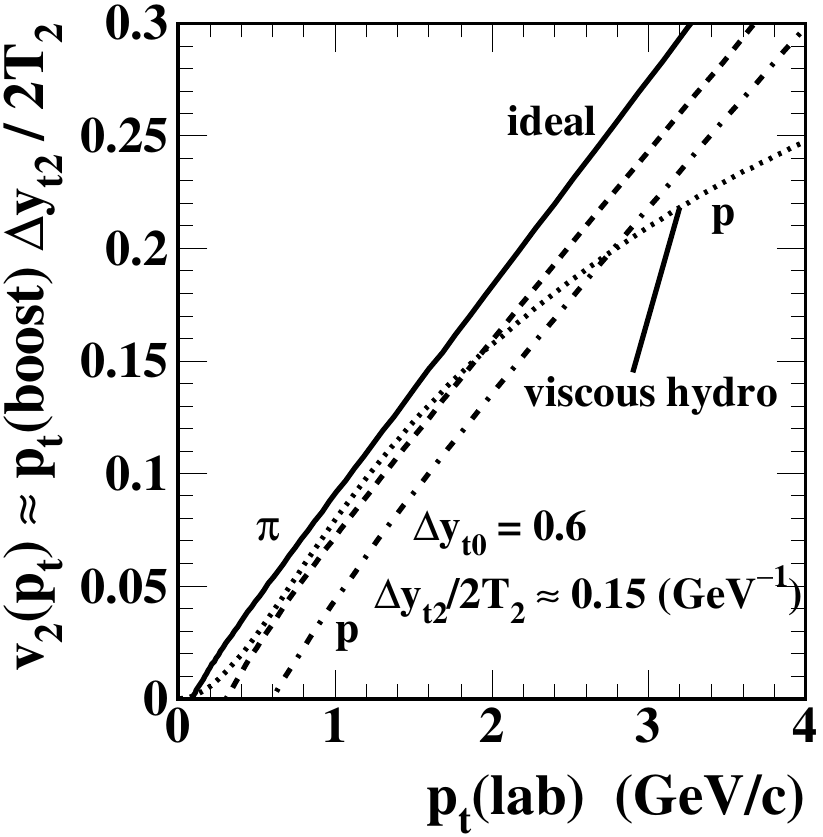}
\put(-70,70) {\bf (c)}
  \includegraphics[width=.24\textwidth]{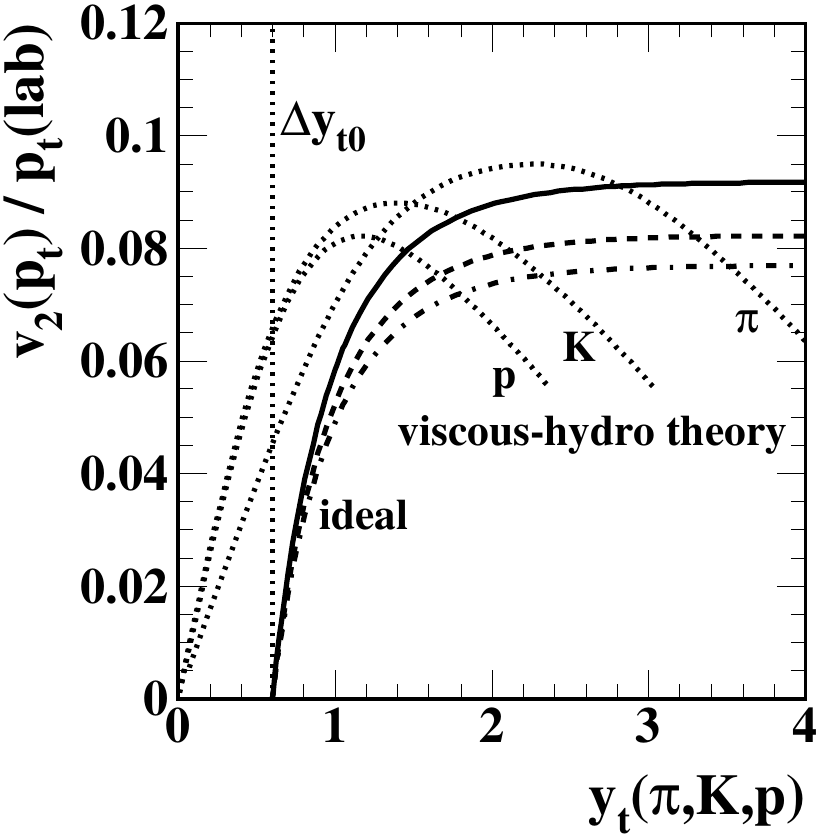}
\put(-20,83) {\bf (d)}
\caption{\label{hydro}
``Ideal-hydro'' kinematic trends assuming single value of source boost $\Delta y_{t0}$ and quadrupole spectrum equal to SP spectrum~\cite{quadspec}. Viscous-hydro theory trends~\cite{rom} are included for comparison.
} 

 \end{figure}

The following simple system illustrates some implications of Eq.~(\ref{v2v2}). The source-boost distribution, which should be broad for Hubble expansion of a bulk medium, is assumed to be a single fixed value $\Delta y_{t0}$ for each collision system. The quadrupole spectrum $\bar \rho_2[y_t,b;\Delta y_{t0}(b)]$ is assumed to coincide with SP spectrum $\bar \rho_0[p_t,b;\Delta y_{t0}(b)]$ including centrality-dependent radial flow~\cite{starblast} measured by $\Delta y_{t0}(b)$. In that case $v_2(p_t) \approx p_t'(\Delta y_{t2} / 2T_2)$ and $v_2(p_t) / p_t \propto p_t' / p_t$. Figure~\ref{hydro} (a) shows $p_t'$ in the boost frame vs lab \pt\ for three hadron species. Figure~\ref{hydro} (b)  shows ratio $p_t' / p_t$ vs \yt\ (with proper hadron mass) as a {\em universal curve} common to all hadron species, with shape determined solely by fixed source boost $\Delta y_{t0} = 0.6$. Figure~\ref{hydro} (c)  shows panel (a) adjusted to anticipate 200 GeV $v_2(p_t)$ data, and panel (d) shows the equivalent for ratio $v_2(p_t) / p_t$. In each case viscous-hydro predictions for 200 GeV \auau\ collisions (dotted curves)~\cite{rom} shown for comparison exhibit striking differences from the ``ideal-hydro'' example. This exercise illustrates ``mass ordering'' at lower \pt\ in the conventional plotting format of panel (a).  The quantitative source-boost distribution that hydro theory actually predicts (broad or narrow?) is more easily tested in the format of panel (d) than in (a) or (c). 

\section{200 GeV Au-Au Quadrupole Spectra} \label{200gevv2}

\begin{figure}[h]
\centering
  \includegraphics[width=.24\textwidth]{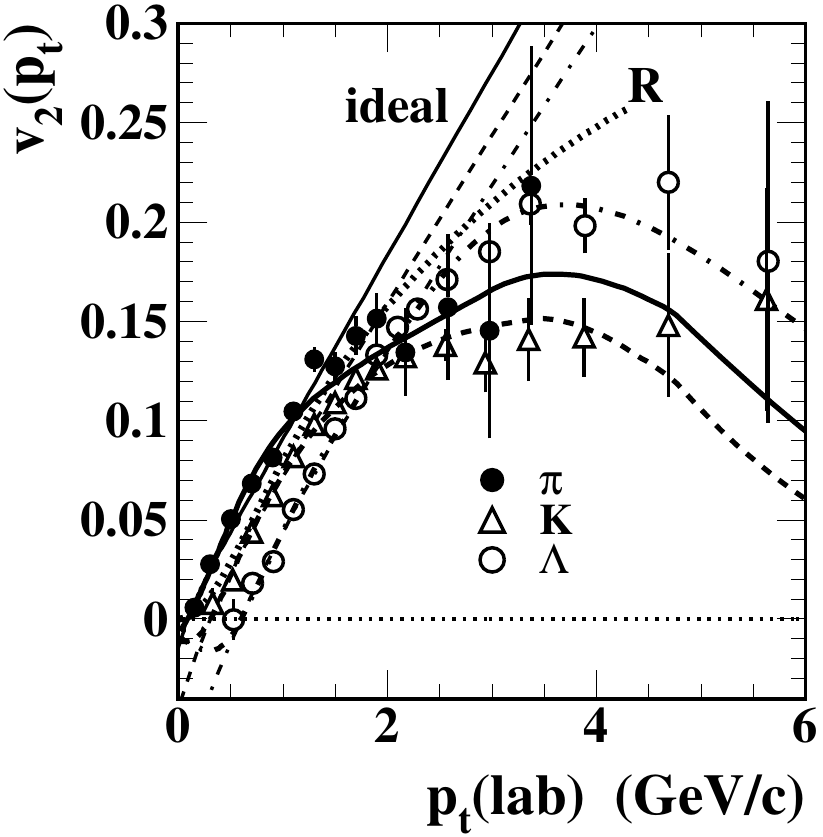}
\put(-72,80) {\bf (a)}
  \includegraphics[width=.24\textwidth]{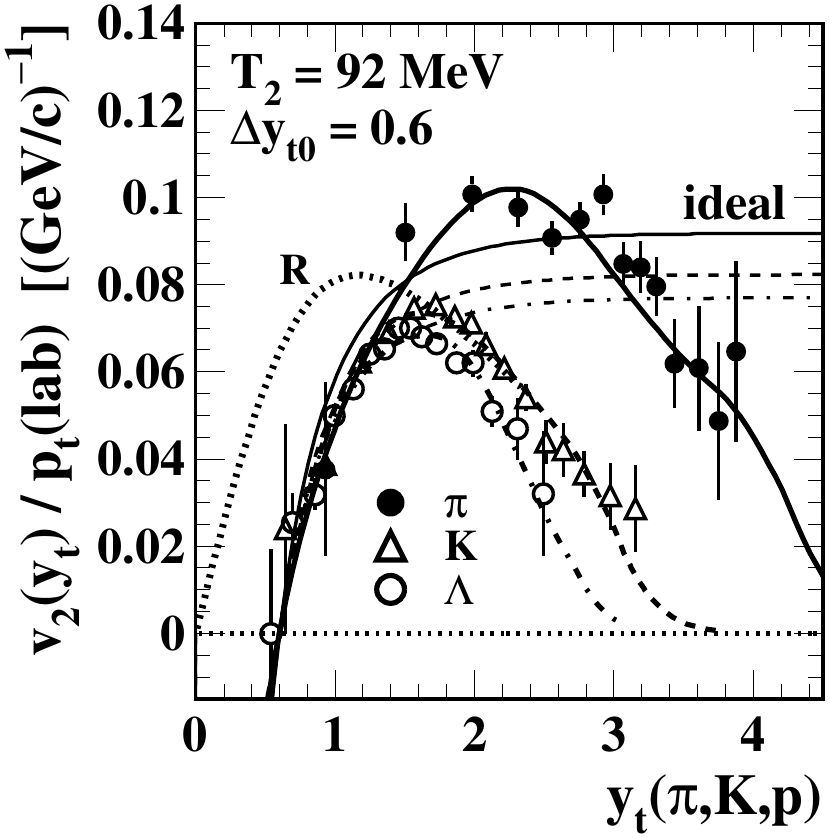}
\put(-20,80) {\bf (b)}
  \includegraphics[width=.24\textwidth]{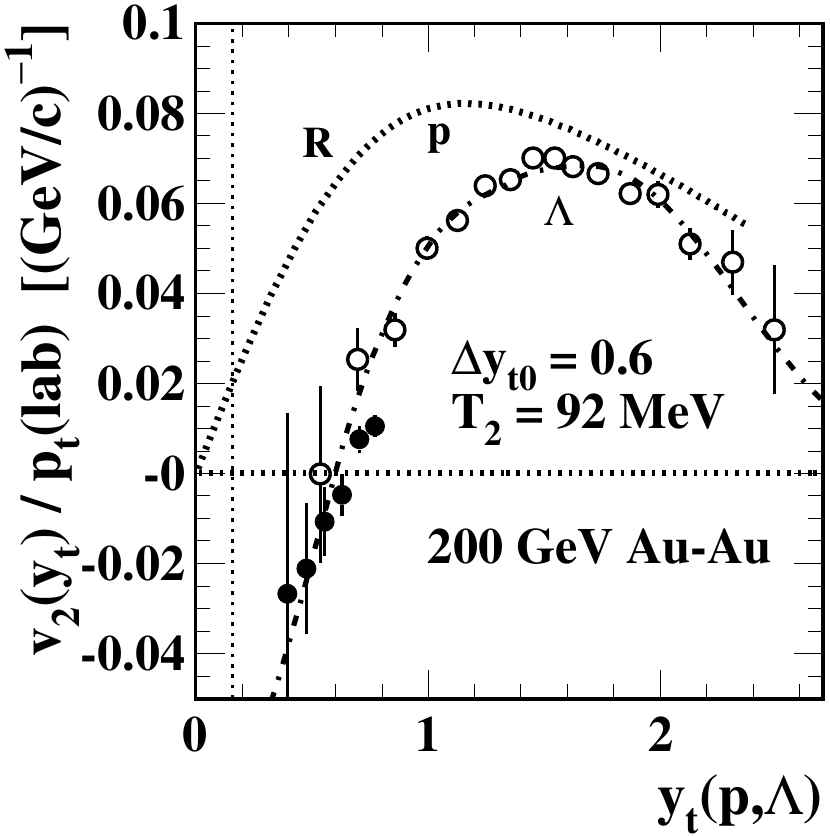}
\put(-20,80) {\bf (c)}
  \includegraphics[width=.24\textwidth]{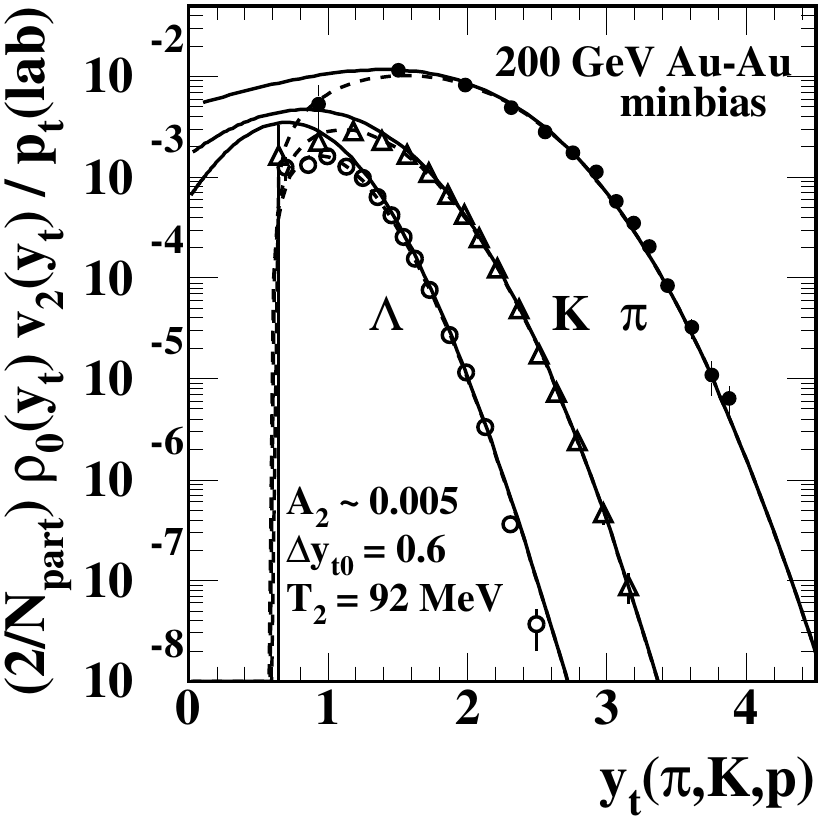}
\put(-20,70) {\bf (d)}
\caption{\label{200gev}
$v_2(p_t)$ data for three hadron species from 0-80\% 200 GeV \auau\ collisions~\cite{v2pions,v2strange} processed to obtain $V_2(y_t) =  \bar \rho_0(y_t) v_2(y_t) \propto p_t'(y_t)\, \bar \rho_2(y_t;\Delta y_{t0})$ quadrupole spectra as in Ref.~\cite{quadspec}.
}  
 \end{figure}

Quadrupole spectra can be inferred from $v_2(p_t,b)$ data in a few steps. Figure~\ref{200gev} (a) shows 0-80\% central $v_2(p_t)$ data for three hadron species in the conventional plotting format~\cite{v2pions,v2strange}. ``Mass ordering'' below 1.5 GeV/c is said to indicate a hydro mechanism. Curves representing ``ideal hydro'' cross the top edge. A viscous-hydro theory curve for protons is indicated by R~\cite{rom}. Curves passing through data at higher \pt\ are explained below. Figure~\ref{200gev} (b) shows the same data in the form $v_2(p_t) / p_t$(lab) vs \yt\ with proper mass for each hadron species. The ideal-hydro curves go to a constant value for higher \yt, and the data trends for three hadron species share a common zero intercept  that can be identified with fixed source boost $\Delta y_{t0} \approx 0.6$. Figure~\ref{200gev} (c) confirms that the (dotted) viscous-hydro theory curve is strongly falsified by Lambda (or proton) data. Figure~\ref{200gev} (d) shows the quadrupole Fourier coefficient as the product $V_2(y_t) =  \bar \rho_0(y_t) v_2(y_t)$ obtained via SP spectra for three  hadron species~\cite{quadspec}. 

\begin{figure}[h]
\centering
\sidecaption
  \includegraphics[width=.5\textwidth]{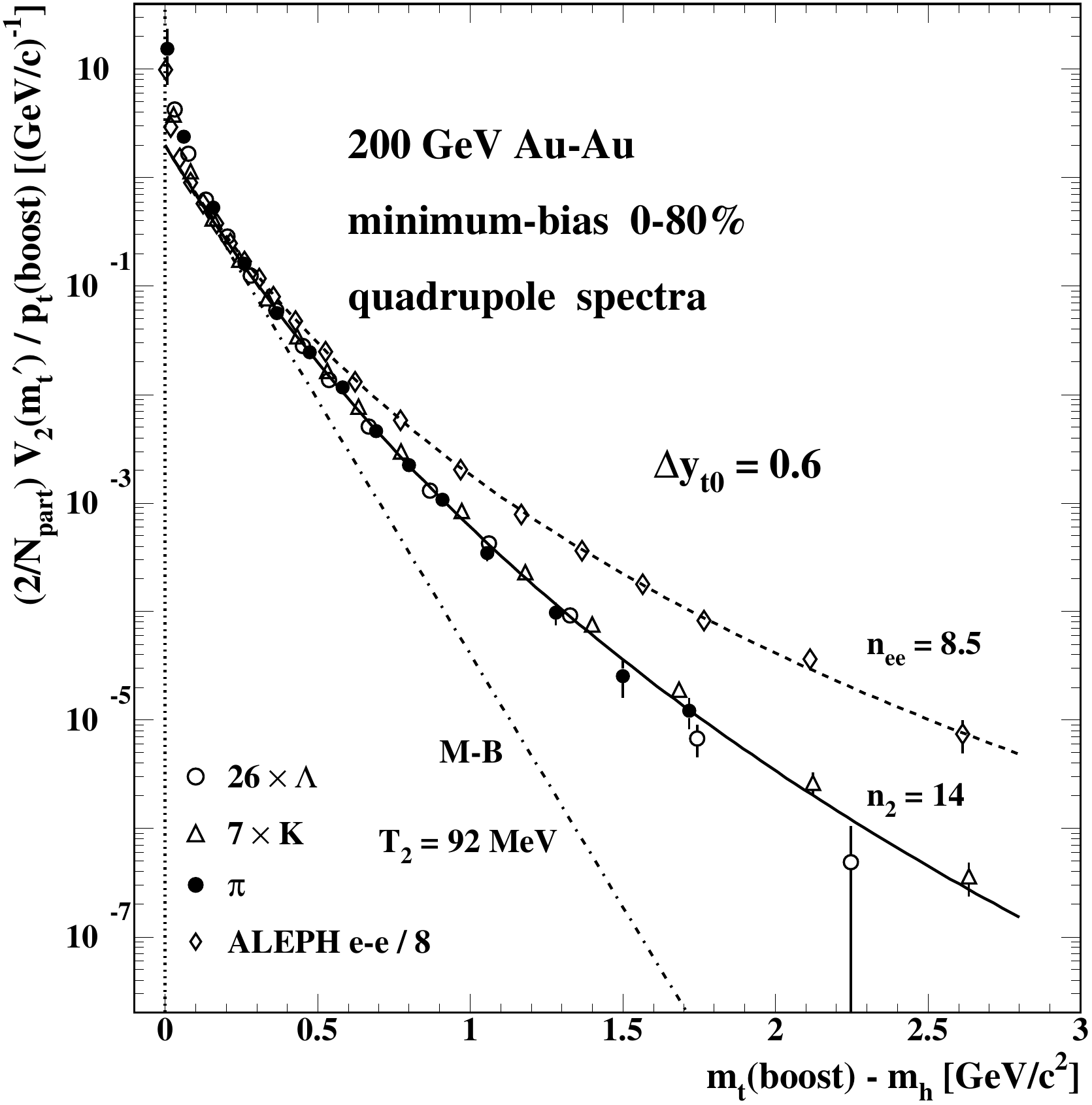}
\caption{\label{quadspec}
Quadrupole spectra from 200 GeV \auau\ collisions for three hadron species (points in legend) rescaled by statistical-model abundances relative to pions (factors noted in the legend) plotted on $m_t'$ in the boost frame. The solid curve is a L\'evy distribution with $T_2 = 92$ MeV (dash-dotted curve) and exponent $n_2 = 14$ very different from \auau\ SP spectrum values $T_0 = 145$ MeV and $n_0 = 12$~\cite{hardspec}. The dashed curve and data are derived from the fragment \pt\ spectrum (perpendicular to the dijet axis) for 91 GeV LEP dijets~\cite{eeptspec} also exhibiting $T \approx 90$ MeV.
}  
 \end{figure}

To  obtain final quadrupole spectra from the data in Fig.~\ref{200gev} (d) three more steps are required. The spectra are multiplied by $p_t(\text{lab}) / p_t(\text{boost})$ determined exactly by inferred $\Delta y_{t0} = 0.6$, shifted left by $\Delta y_{t0}$ from lab frame \yt\ to boost frame $y_t'$ and finally transformed to boost-frame $m_t'$ with the appropriate Jacobian. The result is shown in Figure~\ref{quadspec}. The spectra, rescaled by their statistical-model abundances relative to pions~\cite{statmodel}, lie on a common locus (solid curve) with slope parameter $T_2 = 92$ MeV and L\'evy exponent $n_2 = 14$ dramatically different from the hadron SP spectrum with $T_0 = 145$ MeV and $n_0 = 12$~\cite{hardspec}. The solid curve, back-transformed by reversing the sequence, gives the curves passing through data in Fig.~\ref{200gev}.

\section{2.76 TeV Pb-Pb Quadrupole Spectra} \label{276tevv2}

\begin{figure}[h]
\centering
  \includegraphics[width=.24\textwidth]{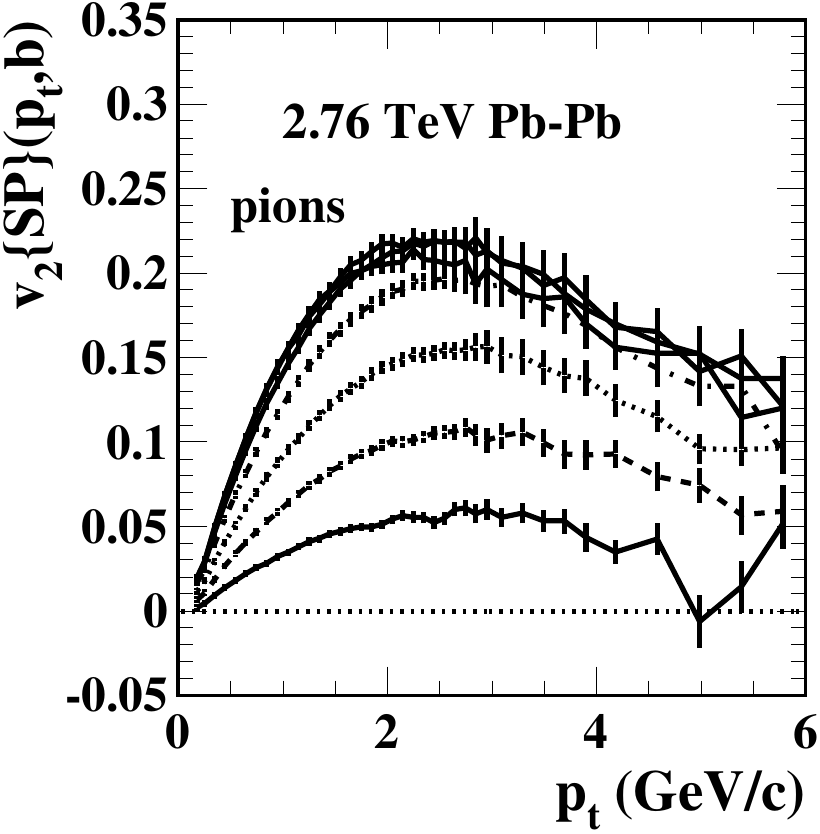}
\put(-20,82) {\bf (a)}
  \includegraphics[width=.24\textwidth]{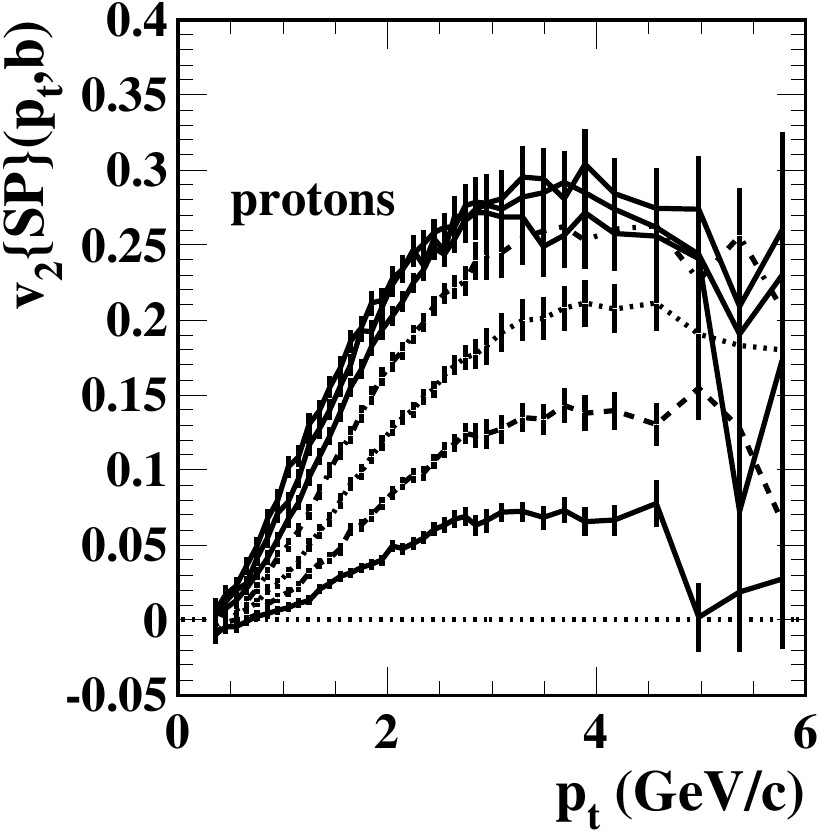}
\put(-20,82) {\bf (b)}
  \includegraphics[width=.24\textwidth]{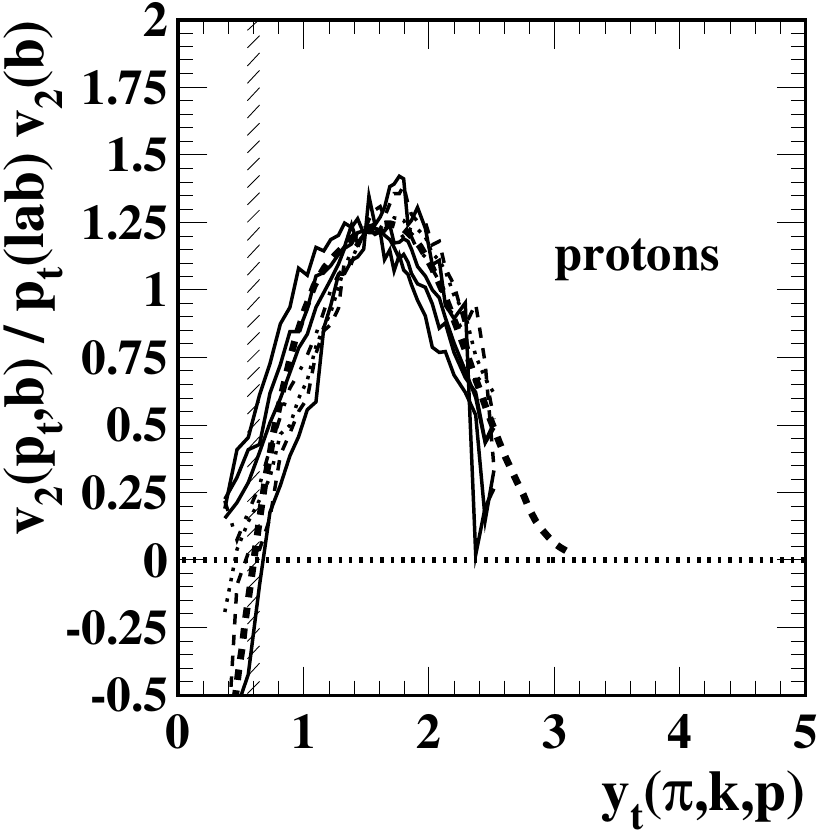}
\put(-20,82) {\bf (c)}
  \includegraphics[width=.24\textwidth]{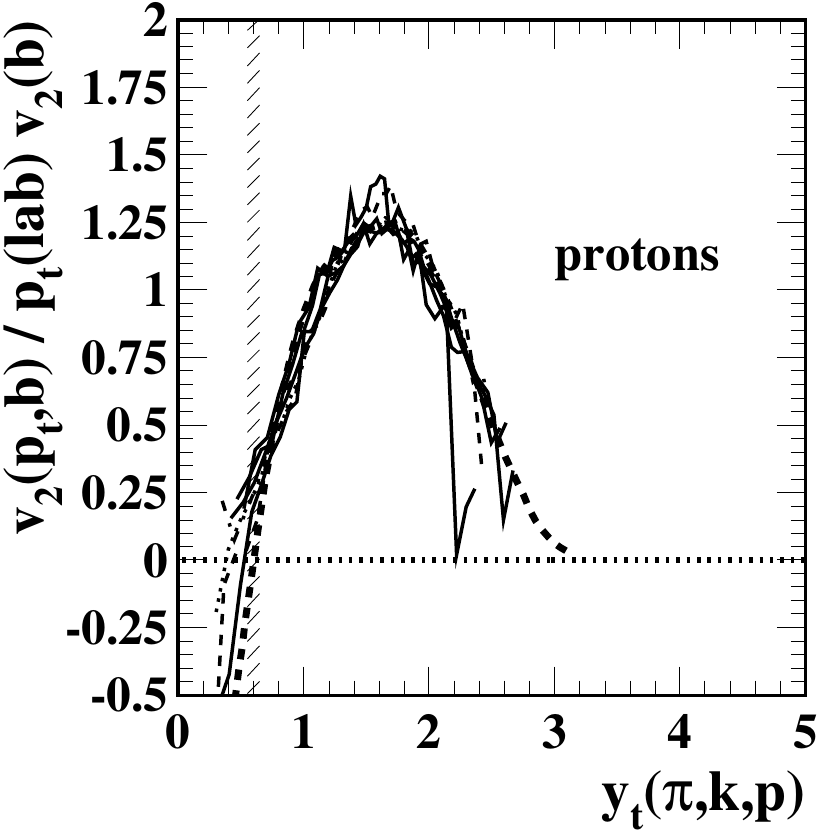}
\put(-20,82) {\bf (d)}
\caption{\label{lhcv2a}
$v_2(p_t,b)$ data vs \pt\ for pions and protons from  2.76 TeV \pbpb\ collisions~\cite{alicev2ptb}.
}  
 \end{figure}

Figure~\ref{lhcv2a} (a,b) shows $v_2(p_t,b)$ data vs \pt\ for pions and protons from seven centralities of 2.76 TeV \pbpb\ collisions~\cite{alicev2ptb}. 200 GeV $v_2(p_t,b)$ data are presented in Ref.~\cite{v2ptb} for comparison. Figure~\ref{lhcv2a} (c) shows proton data plotted as in Fig.~\ref{200gev} (b) but rescaled by \pt-integral values $v_2(b)$~\cite{alicev2b}. The dashed curve is the 200 GeV equivalent. The 2.76 TeV data reveal significant variation of source boost $\Delta y_{t0}(b)$ with centrality. Figure~\ref{lhcv2a} (d) shows the same data with all centralities boosted (shifted on \yt) to coincide with the 30-40\% data. Within uncertainties all centralities follow the same locus that coincides also with the 200 GeV dashed curve.

\begin{figure}[h]
\centering
  \includegraphics[width=.48\textwidth]{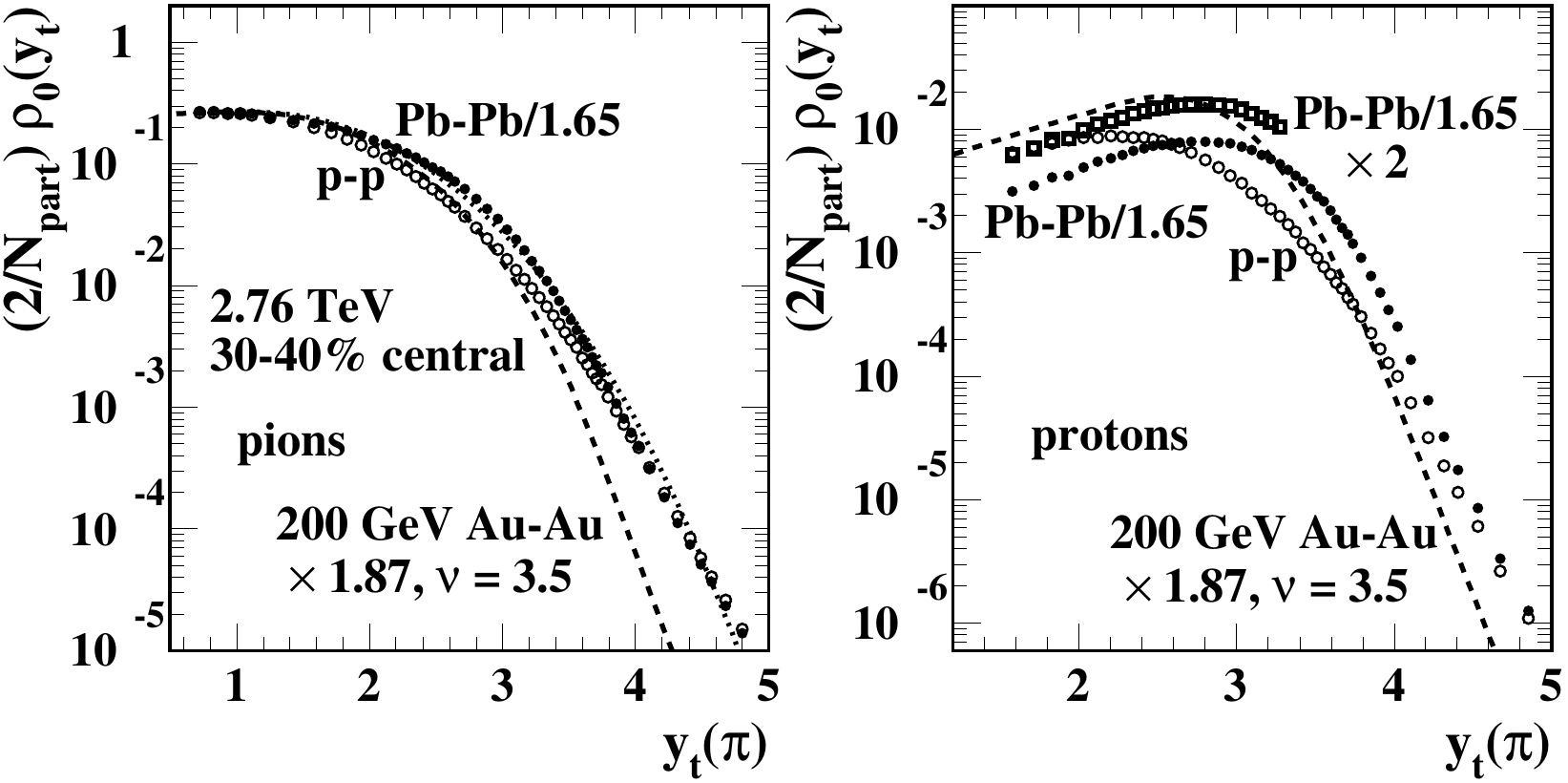}
\put(-20,65) {\bf (b)}
\put(-117,65) {\bf (a)}
  \includegraphics[width=.48\textwidth]{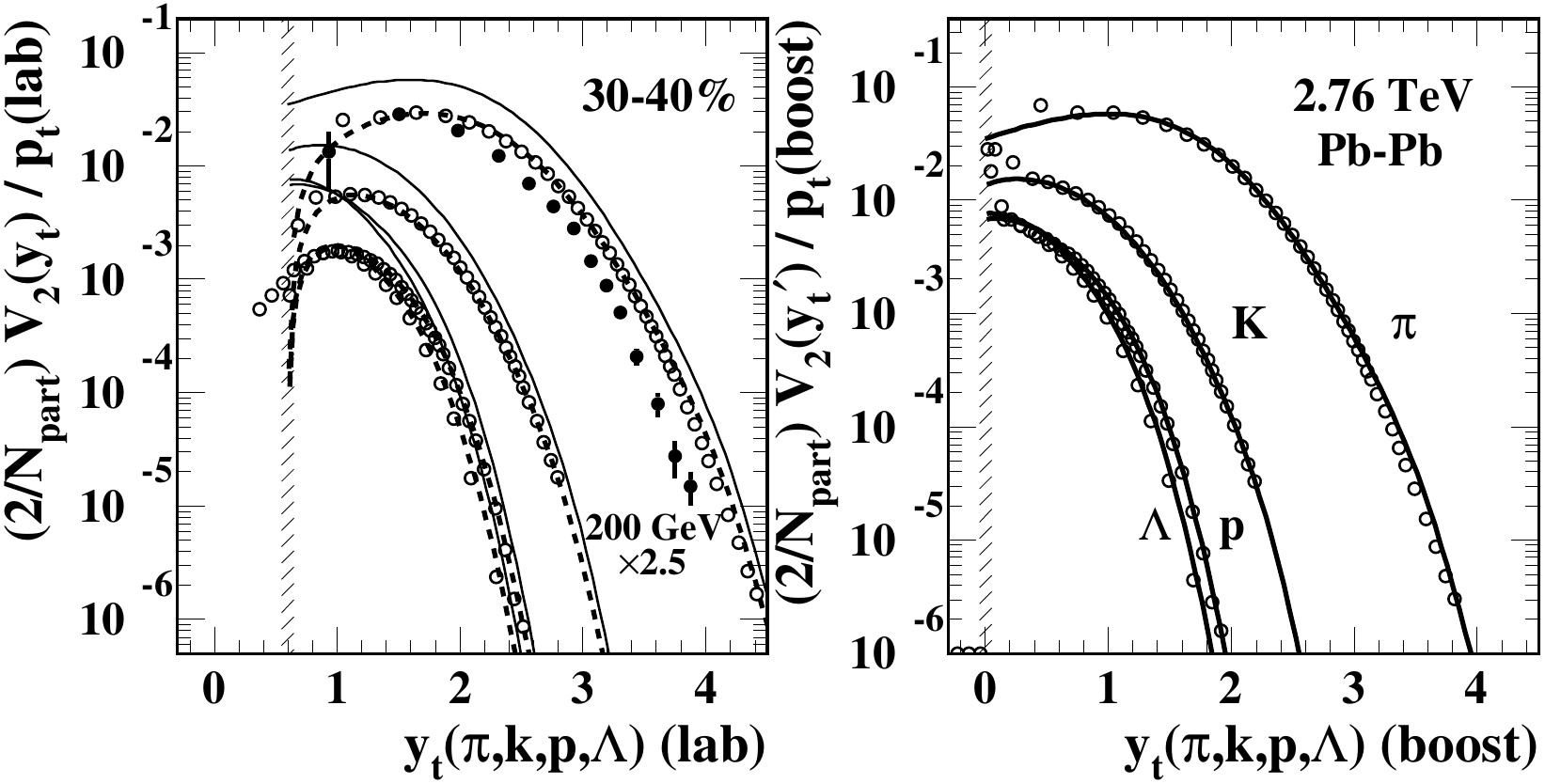}
\put(-18,65) {\bf (d)}
\put(-115,65) {\bf (c)}
\caption{\label{lhcv2b}
SP spectra for pions (a) and protons (b) from 30-40\% central 2.76 TeV \pbpb\ collisions (solid points). Quadrupole spectra for  the same system in the lab frame (c) and boost frame (d).
}  
 \end{figure}

Figure~\ref{lhcv2b} (a,b) shows SP spectra for pions and protons from p-p collisions (open points) and from 30-40\% central 2.76 TeV \pbpb\ (solid dots) as the average of 0-5\% and 60-80\% central collisions with two issues~\cite{alicespec}: (a) The \pbpb\ pion spectrum normalized by $N_{part}/2$ must be divided by factor 1.65 to coincide with the \pp\ spectrum at lower \pt\ in accord with 200 GeV data~\cite{hardspec}. (b) The \pbpb\ proton spectrum normalized by $N_{part}/2$ (solid dots) is divided by the same factor but falls substantially below (factor 2) the \pp\ spectrum at lower \pt\ suggesting significant uncorrected inefficiency in  that interval. Figure~\ref{lhcv2b} (c) shows the equivalent of Fig.~\ref{200gev} (d) for four hadron species. The dashed curves through those data are the dashed curves for 200 GeV $v_2(p_t,b)/p_t(\text{lab})$ multiplied by the \pbpb\ SP spectra rescaled by 1/1.65. Figure~\ref{lhcv2b} (d) shows those data multiplied by factor $p_t(\text{lab}) / p_t(\text{boost})$ and transformed to the boost frame (shifted left by $\Delta y_{t0} = 0.6$).

\begin{figure}[h]
\centering
\sidecaption
  \includegraphics[width=.5\textwidth]{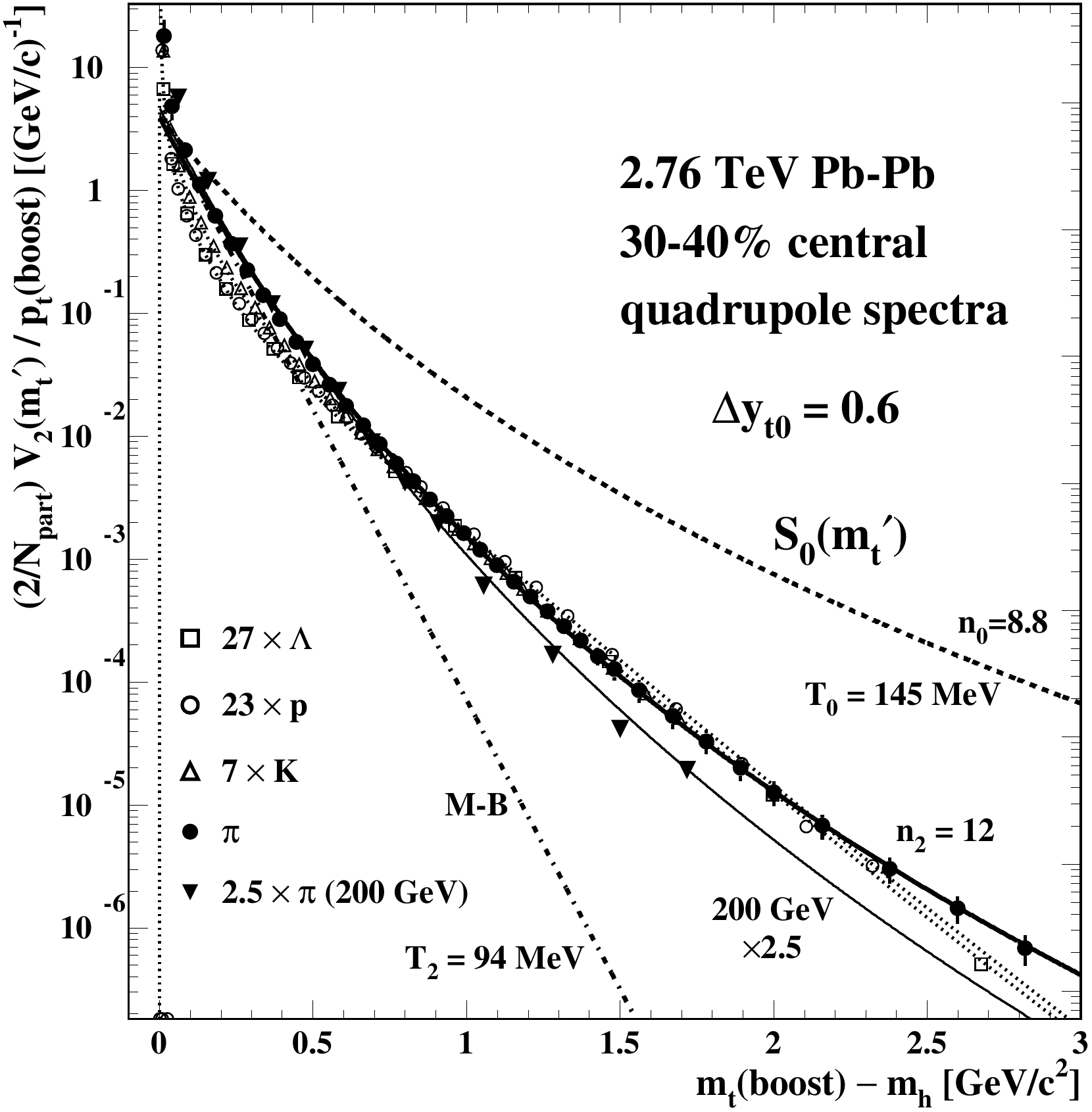}
\caption{\label{lhcquad}
Quadrupole spectra from 2.76 TeV \pbpb\ collisions for four hadron species (points in legend) rescaled by statistical-model abundances relative to pions (factors noted in the legend) plotted on $m_t'$ in the boost frame. The solid curve is a L\'evy distribution with $T_2 = 94$ MeV (dash-dotted curve) and exponent $n_2 = 12$, very different from \pp\ SP spectrum soft component $\hat S_0(m_t')$ (dashed curve) values $T_0 = 145$ MeV and $n_0 = 8.8$~\cite{hardspec}. The 200 GeV quadrupole spectrum multiplied by anticipated energy-scaling factor 2.5  is plotted as the thin solid curve and inverted solid triangles
}  
 \end{figure}

Figure~\ref{lhcquad} shows data and curves in Fig.~\ref{lhcv2b} (d) transformed to $m_t'$ in the boost frame via the proper Jacobian and rescaled by statistical-model abundances relative to pions (points in legend). The pion data fall on the bold solid curve with $T_2 = 94$ MeV and $n_2 = 12$. The 200 GeV quadrupole spectrum data (inverted solid triangles and thin solid curve) are shown multiplied by factor 2.5 expected from observed energy scaling of $v_2$ and $\bar \rho_0$. Slope parameters $T_2$ are not significantly different, but the L\'evy exponent decreases significantly at the higher energy consistent with the trend for SP spectrum soft component $\hat S_0(m_t)$~\cite{alicespec2,alicespec}. Just as at RHIC energies there is a great difference between quadrupole spectra and SP spectra. SP spectrum soft component $\hat S_0(m_t')$ for 2.76 TeV \pp\ collisions plotted on $m_t'$ is shown as the dashed curve. The spectrum for protons (and Lambdas) falls substantially below the bold solid curve for $m_t' < 0.7$ GeV/$c^2$, consistent with the spectrum result in Fig.~\ref{lhcv2b} (b). The dotted curves for kaons and protons are the 200 GeV dashed curves in Fig.~\ref{lhcv2b} (c) derived from the solid curve in Fig.~\ref{quadspec} processed in this case with the corresponding 2.76 TeV SP spectra.

\section{Summary} \label{summ}

Quadrupole spectra for identified hadrons, which may involve only a small fraction of final-state particles that actually ``carry'' the quadrupole component, can be extracted from \pt-differential $v_2(p_t,b)$ data by a simple sequence of transformations given the availability of matching single-particle hadron spectra. The sequence leads to determination of a hadron source-boost distribution consistent with a single value $\Delta y_{t0}$ (for a given collision system) that is inconsistent with the broad distribution expected for Hubble expansion of a flowing bulk medium. Quadrupole spectra at 2.76 TeV are remarkably similar to those for 200 GeV and very different from single-particle hadron spectra, contradicting a basic assumption of the flow narrative that almost all hadrons must participate in such flows. These new results, combined with related trends observed over the past ten years, strongly suggest that the nonjet azimuth quadrupole does not represent a hydrodynamic flow. The NJ quadrupole may instead be the manifestation of a QCD mechanism similar to QED antenna radiation.

\end{document}